\documentclass[aps,pre,10pt,onecolumn,showpacs,superscriptaddress]{revtex4}

\usepackage{times}
\usepackage{amsfonts}
\usepackage{amsmath}
\usepackage{amssymb}
\usepackage{graphicx}

\begin{document}

\title{Energy loss of ions by electric--field fluctuations in a magnetized plasma}
\author{Hrachya B. Nersisyan}
\email{hrachya@irphe.am}
\affiliation{Institute of Radiophysics and Electronics, 0203 Ashtarak, Armenia}
\affiliation{Centre of Strong Fields Physics, Yerevan State University, Alex Manoogian
str. 1, 0025 Yerevan, Armenia}
\author{Claude Deutsch}
\email{claude.deutsch@pgp.u-psud.fr}
\affiliation{LPGP (UMR-CNRS 8578), Universit\'{e} Paris XI, 91405 Orsay, France}
\date{\today }

\begin{abstract}
The results of a theoretical investigation of the energy loss of charged particles in a magnetized
classical plasma due to the electric field fluctuations are reported. The energy loss
for a test particle is calculated through the linear-response theory. At vanishing magnetic field
the electric field fluctuations lead to an energy  gain of the charged particle for all velocities.
It has been shown that in the presence of strong magnetic field this effect occurs only at low--velocities.
In the opposite case of high-velocities the test particle systematically loses its energy due to the
interaction with a stochastic electric field. The net effect of the fluctuations is the systematic
reduction of the total energy loss (i.e. the sum of the polarization and stochastic energy losses)
at vanishing magnetic field and reduction or enhancement at strong field depending on the velocity
of the particle. It is found that the energy loss of the slow heavy ion contains an anomalous term
which depends logarithmically on the projectile mass. The physical origin of this anomalous term is
the coupling between the cyclotron motion of the plasma electrons and the long--wavelength, low--frequency
fluctuations produced by the projectile ion. This effect may strongly enhance the stochastic energy
gain of the particle.
\end{abstract}

\pacs{52.40.Mj, 52.25.Xz, 52.25.Gj, 52.25.Fi}
\maketitle

\section{Introduction}
\label{sec:intr}

There is an ongoing interest in the theory of interaction of charged particle beams with plasmas. Although
most theoretical works have reported on the energy loss of ions in a plasma without magnetic field, the
strongly magnetized case has not yet received as much attention as the field-free case. The energy loss of
ion beams and the related processes in magnetized plasmas are important in many areas of physics such as
transport, heating, magnetic confinement of thermonuclear plasmas, and astrophysics. Recent applications are
the cooling of heavy-ion beams by electrons \cite{sor83,ner07,mol03} and the energy transfer for magnetized
target fusion \cite{cer00}, or heavy-ion inertial confinement fusion.

For a theoretical description of the energy loss of ions in a plasma, there exist some standard approaches.
The dielectric linear response (LR) treatment considers the ion as a perturbation of the target plasma and
the stopping is caused by the polarization of the surrounding medium. It is generally valid if the ion couples
weakly to the target. Since the early 1960s, a number of calculations of the stopping power within
LR treatment in a magnetized plasma have been presented (see
Refs.~\cite{ros60,akh61,hon63,may70,ner98,ner00,ner03,ner11,see98,wal00,ste01,deu08}
and references therein). Alternatively, the stopping is calculated as a result of the energy transfers in
successive binary collisions (BCs) between the ion and the electrons \cite{ner09,zwi99,zwi02,toe02}. Here it
is necessary to consider appropriate approximations for the screening of the Coulomb potential by the plasma
\cite{ner07}. However, significant gaps between these approaches involve the crucial ion stopping along magnetic
field $\mathbf{B}$ and perpendicular to it. In particular, at high $B$ values, the BC predicts a vanishingly
parallel energy loss, which remains at variance with the nonzero LR one. Also challenging BC-LR discrepancies
persist in the transverse direction, especially for vanishingly small ion projectile velocity $v$ when the
friction coefficient contains an anomalous term diverging logarithmically at $v\to 0$ \cite{ner00,ner03}.
For calculation of the energy loss of an ion two new alternative approaches have been recently suggested. One
of these methods is specifically aimed at a low-velocity energy loss, which is expressed in terms of
velocity--velocity correlation and, hence, to a diffusion coefficient \cite{deu08}. Next in Ref.~\cite{ner11}
using the Bhatnagar-Krook-Gross approach based on the Boltzmann-Poisson equations for a collisional and
magnetized classical plasma the energy loss of an ion is studied through a LR approach, which is constructed
such that it conserves particle local number.

Previous treatments of the energy loss of charged particles in a magnetized plasma
\cite{ner07,mol03,cer00,ros60,akh61,hon63,may70,ner98,ner00,ner03,ner11,see98,wal00,ste01,deu08} were obtained
by treating the plasma in an average manner, i.e., microscopic fluctuations were neglected. However, any
media and, in particular, any plasma being a statistical system, is characterized by stochastic fluctuations.
These fluctuations couple to external perturbations and the response of the medium to these perturbations can
be expressed through suitable correlation functions of the microscopically fluctuating variables. It is well
known that the motion of charged particles in such an environment is stochastic in nature and resembles Brownian
motion. The effect of the electromagnetic field fluctuations on the energy loss of a charged particle moving in
an unmagnetized classical plasma has been studied by several authors in the past \cite{gas65,hub61,bek66,sit67,akh75,kal61,tho60,ich73}.
This effect leads to an energy gain of the particles and is most effective in the low--velocity limit. In addition
the effect of the fluctuations are mainly important for light particles (e.g., for an electron) since this effect
is inversely proportional to the mass of the particle and is negligible for heavy ions. Recently similar problem
has been formulated for the energy loss in a so--called quark-gluon plasma \cite{cha07} which is expected to be
formed in relativistic heavy--ion collision experiments.
Since the subject of the energy loss is of topical interest, it is the principal motivation of the present paper to
quantitatively investigate the effect of microscopic fluctuations on the energy loss of a charged particle passing
through an equilibrium, weakly--coupled magnetized plasma. In particular, our main objective is to study through the
LR theory the effect of a strong magnetic field on the energy loss process arising due to the fluctuations in a
plasma. The sequel is structured as follows. Interactions of a gyrating projectile with target stochastic electric
fields are investigated at length in Sec.~\ref{sec:2}. Resulting energy loss rates are respectively explained in
Sec.~\ref{sec:3} for a target plasma with no magnetic field and then with an infinite one as well as in Sec.~\ref{sec:4}
for a finite magnetic field and at low--velocity of the particle.

\section{Interaction of a gyrating particle with plasma stochastic electric field}
\label{sec:2}

The dielectric function approach was adopted earlier to calculate the energy loss of charged particles due to polarization
effects of the medium \cite{ner07,mol03,cer00,ros60,akh61,hon63,may70,ner98,ner00,ner03,ner11,see98,wal00,ste01,deu08}.
It is assumed that the energy lost by the particle is small compared to the energy of the particle
itself so that the change in the velocity of the particle during the motion may be neglected, i.e, the particle
moves in a straight-line (in $B= 0$ case) or helical (in $B\neq 0$ case) trajectory. We consider a nonrelativistic projectile
particle with charge $Ze$ ($-e$ is the electron charge) and with a velocity $\mathbf{v}(t)$, which moves in a magnetized
classical plasma with a constant
magnetic field $\mathbf{B}$. We ignore any role of the electron spin or magnetic moment due to the nonrelativistic motion
of the particle and the plasma electrons. The energy loss of a particle is determined by the work of the retarding forces
acting on the particle in the plasma from the electric field generated by the particle itself while moving. So the energy
loss of the particle per unit time, i.e. energy loss rate (ELR), is given by
\cite{ner07,mol03,cer00,ros60,akh61,hon63,may70,ner98,ner00,ner03,ner11,see98,wal00,ste01,deu08}
\begin{equation}
\dfrac{dW}{dt}=Ze\mathbf{v}(t)\cdot \mathbf{E}\left( \mathbf{r}(t),t\right) ,
\label{eq:1}
\end{equation}%
where the electric field is taken at the location $\mathbf{r}(t)$ of the test particle. Within linear response theory
the correlation function of the fluctuations of charge and current densities and the electromagnetic fields in the medium
with space-time dispersion are completely determined in terms of the dielectric tensor of the medium. The total electric
field $\mathbf{E}(\mathbf{x},t)$ induced in the plasma can be related to the external charge of the test particle by
solving Poisson's equation (see, e.g., Ref.~\cite{ich73})
\begin{equation}
\mathbf{E}\left( \mathbf{x},t\right) =-\frac{2iZe}{\left( 2\pi \right) ^{3}}%
\int_{-\infty }^{\infty }d\omega \int d\mathbf{k}e^{i\mathbf{k}\cdot \mathbf{%
x}-i\omega t}\frac{\mathbf{k}}{k^{2}\varepsilon \left( \mathbf{k},\omega
\right) }\int_{-\infty }^{\infty }dt^{\prime }e^{i\omega t^{\prime }-i%
\mathbf{k}\cdot \mathbf{r}(t^{\prime })} .
\label{eq:2}
\end{equation}%
Here $\varepsilon (\mathbf{k},\omega )=(k_{i}k_{j}/k^{2})\varepsilon _{ij}(\mathbf{k},\omega )$ is the longitudinal dielectric
function and $\varepsilon _{ij}(\mathbf{k},\omega )$ is the dielectric tensor of the plasma, respectively.

The previous formula for the energy loss in Eq.~\eqref{eq:1} does not take into account the fluctuations of the electric fields
in the plasma and the particle recoil in interactions. To accommodate these effects it is necessary to replace Eq.~\eqref{eq:1}
with \cite{sit67,akh75},
\begin{equation}
\dfrac{dW}{dt}=Ze\left\langle \mathbf{v}\left( t\right) \cdot \boldsymbol{%
\mathcal{E}}\left( \mathbf{r}\left( t\right) ,t\right) \right\rangle ,
\label{eq:3}
\end{equation}%
where $\langle \cdots \rangle $ denotes the statistical averaging operation.
It is to be noted that here two averaging procedures are performed: (i) an
ensemble average with respect to the equilibrium distribution function and (ii) a
time average over random fluctuations in plasma. These two operations are
commuting and only after both of them are performed the average quantity
takes up a smooth value \cite{kal61}. In the following, we will explicitly
denote the ensemble average by $\langle \cdots \rangle _{a}$ wherever
required to avoid possible confusion.

The electric field $\boldsymbol{\mathcal{E}}(\mathbf{x},t)=\mathbf{E}(%
\mathbf{x},t)+\tilde{\mathbf{E}}(\mathbf{x},t)$ in Eq.~\eqref{eq:3} consists
of the induced field $\mathbf{E}(\mathbf{x},t)$ given by Eq.~\eqref{eq:2}
and a spontaneously generated microscopic field $\tilde{\mathbf{E}}(\mathbf{x%
},t)$, the latter being a random function of position and time.

The classical equation of motion of the particle with charge $Ze$ and mass $M $ moving in
the fluctuating electric field $\boldsymbol{\mathcal{E}}(\mathbf{x},t)$ and a uniform
magnetic field $\mathbf{B}$ have the form
\begin{equation}
\dot{\mathbf{v}}(t)=\varsigma \Omega _{c}\left[ \mathbf{v}\left( t\right)
\times \mathbf{b}\right] +\frac{Ze}{M}\boldsymbol{\mathcal{E}}\left( \mathbf{%
r}(t),t\right) ,
\label{eq:4}
\end{equation}%
where $\varsigma =\vert Z\vert /Z$, $\Omega _{c}=\vert Z\vert eB/Mc$ is the cyclotron frequency of the
test particle, $\mathbf{b}=\mathbf{B}/B$ is the unit vector along the (external) magnetic field, $%
\boldsymbol{\mathcal{E}}(\mathbf{x},t)$ is the total electric field involved in Eq.~\eqref{eq:3},
$\mathbf{r}(t)$ is the coordinate of the particle. Note that Eq.~\eqref{eq:4} is the generalization
of the Langevin equation in the presence of a magnetic field where the role of the friction force
plays the polarization force determined by a regular electric field $\mathbf{E}(\mathbf{x},t)$ given
by Eq.~\eqref{eq:2}.

Introducing the cyclotron rotation matrix $\mathcal{L}_{ij}(t)$ the equation of motion \eqref{eq:4}
has a formal solution
\begin{eqnarray}
v_{i}\left( t\right) =V_{i}\left( t\right) +\frac{Ze}{M}\int_{-\infty
}^{t}dt^{\prime }\mathcal{R}_{ij}\left( t-t^{\prime }\right) \mathcal{E}%
_{j}\left( \mathbf{r}(t^{\prime }),t^{\prime }\right) ,  \label{eq:5} \\
x_{i}\left( t\right) =X_{i}\left( t\right) +\frac{Ze}{M\Omega _{c}}%
\int_{-\infty }^{t}dt^{\prime }\mathcal{L}_{ij}\left( t-t^{\prime }\right)
\mathcal{E}_{j}\left( \mathbf{r}(t^{\prime }),t^{\prime }\right) , \label{eq:6}
\end{eqnarray}%
where $\mathcal{R}_{ij}(t)=(1/\Omega _{c})d\mathcal{L}_{ij}(t)/dt$, $\mathbf{%
R}(t)$ and $\mathbf{V}(t)$ with the components $X_{i}(t)$ and $V_{i}(t)$,
respectively, are the initial (at $t=-\infty $) unperturbed coordinate and
velocity of the particle free cyclotron motion, respectively
\begin{eqnarray}
&&X_{i}\left( t\right) =x_{0i}+\frac{1}{\Omega _{c}}\mathcal{L}_{ij}\left(
t\right) v_{0j},  \label{eq:7} \\
&&V_{i}\left( t\right) =\mathcal{R}_{ij}\left( t\right) v_{0j} .  \nonumber
\end{eqnarray}%
Here the variables $\mathbf{v}_{0}$ and $\mathbf{r}_{0}$ with the components $v_{0i}$ and $x_{0i}$,
respectively, are independent and are determined by initial conditions for the free cyclotron motion.
Also
\begin{equation}
\mathcal{L}_{ij}\left( t\right) =b_{i}b_{j}\Omega _{c}t+\left( \delta
_{ij}-b_{i}b_{j}\right) \sin \left( \Omega _{c}t\right) +\varsigma
e_{ijs}b_{s}\left[ 1-\cos \left( \Omega _{c}t\right) \right] ,
\label{eq:8}
\end{equation}%
where $b_{i}$ is the component of the unit vector $\mathbf{b}$ and $e_{ijs}$ is the fully antisymmetrical
unit tensor of rank three.

The mean change of the energy of the particle per unit time is given by Eq.~\eqref{eq:3}. The stochastic
time dependence as embodied in Eq.~\eqref{eq:3} comes because of explicit fluctuations in time and motion
of the particle from one field point to another. The dependence on the latter will be expanded about the
free cyclotron motion where the former will be left untouched. This is done by taking a time interval
$\Delta t$ sufficiently large with respect to the time scale of random electric field fluctuations in the
plasma ($\tau _{2}$) but small compared with the time during which the particle motion changes appreciably
($\tau _{1}$), $\tau _{2}\ll \Delta t\ll \tau _{1}$. Explicitly, $\tau _{1}\sim [\epsilon^{2}T^{3/2}\ln%
(1/\epsilon )]^{-1}$ and $\tau _{2}\sim (\epsilon T^{3/2})^{-1}$ \cite{kal61}, where $\epsilon =(4\pi n_{e}%
{\lambda }_{\mathrm{D}}^{3})^{-1}$ is the plasma parameter ($\lambda _{\mathrm{D}}=(T/4\pi n_{e}e^{2})^{1/2}$
and $n_{e}$ are the Debye screening length and electron density, respectively) and $T $ is the plasma
temperature measured in energy units. Of course, in the present context of a magnetized plasma the time
scales $\tau_{1}$ and $\tau_{2}$ may be strongly modified by the magnetic field, see, e.g., Ref.~\cite{ner11}.
In fact in the Coulomb logarithm determining the time $\tau_{1}$, the Debye length is replaced by the cyclotron
radius~\cite{mon74}. This results in a logarithmic dependence on magnetic field in the relaxation time
$\tau_{1}$. On the other hand, at strong magnetic fields $\tau_{2}$ is replaced by the cyclotron period
$\tau_{2}\sim \omega^{-1}_{c}$, where $\omega_{c}= eB/mc$ is the cyclotron frequency of the electrons.
Thus, independently of the strength of the magnetic field and in the weak coupling limit with $\epsilon \ll 1$,
these scales are widely separated. In Eqs.~\eqref{eq:5} and \eqref{eq:6} keeping only the leading order
terms in the expansion we obtain
\begin{equation}
v_{i}\left( t\right) =V_{i}\left( t\right) +\frac{Ze}{M}\int_{-\infty
}^{t}dt^{\prime }\mathcal{R}_{ij}\left( t-t^{\prime }\right) \mathcal{E}%
_{j}\left( t^{\prime }\right) ,
\label{eq:9}
\end{equation}%
\begin{equation}
\mathcal{E}_{i}\left( \mathbf{r}(t),t\right) =\mathcal{E}_{i}\left( t\right)
+\frac{Ze}{M\Omega _{c}}\frac{\partial \mathcal{E}_{i}\left( t\right) }{%
\partial X_{j}}\int_{-\infty }^{t}dt^{\prime }\mathcal{L}_{js}\left(
t-t^{\prime }\right) \mathcal{E}_{s}\left( t^{\prime }\right) ,
\label{eq:10}
\end{equation}%
where $\boldsymbol{\mathcal{E}}(t)\equiv \boldsymbol{\mathcal{E}}(\mathbf{R}%
(t),t)$ and $\mathbf{R}(t)$ is determined by Eq.~\eqref{eq:7}. Substituting
Eqs.~\eqref{eq:9} and \eqref{eq:10} into Eq.~\eqref{eq:3} and neglecting the
third order correlations we obtain
\begin{equation}
\dfrac{dW}{dt}=\dfrac{dW_{0}}{dt}+\dfrac{dW_{\mathrm{st}}}{dt},
\label{eq:11}
\end{equation}%
with
\begin{equation}
\dfrac{dW_{0}}{dt}=Ze\mathbf{V}(t)\cdot \left\langle \boldsymbol{\mathcal{E}}%
\left( t\right) \right\rangle ,
\label{eq:12}
\end{equation}%
\begin{equation}
\dfrac{dW_{\mathrm{st}}}{dt}=\frac{Z^{2}e^{2}}{M}\left\{ \int_{-\infty
}^{t}dt^{\prime }\mathcal{R}_{ij}\left( t-t^{\prime }\right) \left\langle
\mathcal{E}_{i}\left( t\right) \mathcal{E}_{j}\left( t^{\prime }\right)
\right\rangle _{a}+\frac{V_{i}(t)}{\Omega _{c}}\int_{-\infty }^{t}dt^{\prime
}\mathcal{L}_{js}\left( t-t^{\prime }\right) \left\langle \frac{\partial
\mathcal{E}_{i}\left( t\right) }{\partial X_{j}}\mathcal{E}_{s}\left(
t^{\prime }\right) \right\rangle _{a}\right\} .
\label{eq:13}
\end{equation}%
Because the mean value of the fluctuating part of the electric field equals zero, $%
\langle \tilde{\mathbf{E}}(\mathbf{x},t)\rangle =0$, $\langle \boldsymbol{%
\mathcal{E}}\left( t\right) \rangle $ equals the electric field $\mathbf{E}(%
\mathbf{R}(t),t)$ produced by the particle itself in the plasma. Therefore
the first term in Eq.~\eqref{eq:11} determined by Eq.~\eqref{eq:12}
corresponds to the usual polarization energy loss of the particle calculated for
instance in Ref.~\cite{ner98} for a magnetized plasma. For a gyrating projectile particle Eq.~%
\eqref{eq:12} has been evaluated explicitly in Ref.~\cite{ner98} (see also Ref.~\cite{ste01}). Averaging
the energy loss $dW_{0}/dt$ with respect to the test particle cyclotron period $2\pi
/\Omega _{c}$ (we denote this quantity as $\langle dW_{0}/dt\rangle \equiv
S_{0}$) we obtain (see Ref.~\cite{ner98} for details)
\begin{equation}
S_{0}=\frac{Z^{2}e^{2}}{2\pi ^{2}}\sum_{n=-\infty }^{\infty }\int d\mathbf{k}%
\frac{\zeta _{n}\left( \mathbf{k}\right) }{k^{2}}J_{n}^{2}(k_{\perp }a)%
\mathrm{Im}\frac{1}{\varepsilon \left( \mathbf{k},\zeta _{n}\left( \mathbf{k}%
\right) \right) } ,
\label{eq:14}
\end{equation}%
where $\zeta _{n}(\mathbf{k})=k_{\parallel }v_{0\parallel }+n\Omega _{c}$, $%
k_{\parallel }=(\mathbf{k}\cdot \mathbf{b})$, $k_{\perp
}^{2}=k^{2}-k_{\parallel }^{2}$, $J_{n}$ is the Bessel function of the $n$th
order. Here $v_{0\parallel }$ and $v_{0\perp }\geqslant 0$ are the
unperturbed test particle velocities parallel and transverse to the magnetic field
direction $\mathbf{b}$, respectively, and $a=v_{0\perp }/\Omega _{c}$ is the
cyclotron radius.

Assuming that there exists a hierarchy of scales $\tau _{2}\ll \Delta t\ll \tau_{1}$, it can be shown that the
polarization field does not contribute to leading order in the correlations functions appearing in the second
term in Eq.~\eqref{eq:11} as in the case of higher-order Fokker-Planck coefficients \cite{hub61,ich73}. These
terms correspond to the statistical change in the energy of the projectile particle in the plasma due to the
fluctuations of the electric field as well as the velocity of the particle under the influence of this field.
The first term in Eq.~\eqref{eq:13} corresponds to the statistical part of the dynamic friction due to the
space-time correlation in the fluctuations in the electric field, whereas the second one corresponds to the
average change in the energy of the moving particle due to the correlation between the fluctuation in the velocity
of the particle and the fluctuation in the electrical field in the plasma. The temporal averaging in Eq.~\eqref{eq:13},
by definition, includes many random fluctuations over the mean cyclotron motion. However, the correlation function
of these fluctuations are suppressed beyond their characteristic time scales. This allows us to formally extend
the limits in time integrations in Eq.~\eqref{eq:13} to infinity.

Consider now the energy loss rate \eqref{eq:13} of the particle due to the interaction with stochastic plasma electric
field. We consider homogeneous plasma and only electrostatic electric fields. In this case the correlation function
$\langle \mathcal{E}_{i}(\mathbf{x},t)\mathcal{E}_{j}(\mathbf{x}^{\prime },t^{\prime })\rangle _{a}$ in Eq.~\eqref{eq:13}
is a function of $\mathbf{x}-\mathbf{x}^{\prime }$ and $t-t^{\prime }$. Introducing the power spectrum $P(\mathbf{k},\omega )$
(which is determined by $\langle \mathcal{E} ^{2}(\mathbf{x},t)\rangle _{a}$) of the fluctuating electric field this
correlation function reads \cite{sit67,akh75}
\begin{equation}
\left\langle \mathcal{E}_{i}\left( \mathbf{x},t\right) \mathcal{E}_{j}\left(
\mathbf{x}^{\prime },t^{\prime }\right) \right\rangle _{a}=\int d\mathbf{k}%
\int_{-\infty }^{\infty }d\omega e^{i\mathbf{k}\cdot (\mathbf{x}-\mathbf{x}%
^{\prime })-i\omega (t-t^{\prime })}\frac{k_{i}k_{j}}{k^{2}}P\left( \mathbf{k%
},\omega \right) .
\label{eq:15}
\end{equation}%
Similar expression can be obtained for the correlation function involved in the second term of Eq.~\eqref{eq:13}
and containing the field gradient. Substituting these expressions into Eq.~\eqref{eq:13} we arrive at
\begin{equation}
\dfrac{dW_{\mathrm{st}}}{dt}=\frac{Z^{2}e^{2}}{M\Omega _{c}}\int \frac{d%
\mathbf{k}}{k^{2}}\int_{-\infty }^{\infty }d\omega P\left( \mathbf{k},\omega
\right) \left( \frac{\partial }{\partial t}+i\omega \right) H\left( \mathbf{k%
},\omega ,t\right) ,
\label{eq:16}
\end{equation}%
where
\begin{equation}
H\left( \mathbf{k},\omega ,t\right) =\int_{-\infty }^{t}dt^{\prime }\mathcal{%
L}\left( t-t^{\prime }\right) e^{i\mathbf{k}\cdot (\mathbf{R}(t)-\mathbf{R}%
(t^{\prime }))-i\omega (t-t^{\prime })} ,
\label{eq:17}
\end{equation}%
\begin{equation}
\mathcal{L}\left( t\right) =k_{i}k_{j}\mathcal{L}_{ij}\left( t\right)
=k_{\parallel }^{2}\Omega _{c}t+k_{\perp }^{2}\sin \left( \Omega
_{c}t\right) .
\label{eq:18}
\end{equation}

For evaluation of the function $H(\mathbf{k},\omega ,t)$ we substitute the
unperturbed coordinate (see Eq.~\eqref{eq:7}) of the free helical motion of
the particle into Eq.~\eqref{eq:17}. The time-integration in Eq.~\eqref{eq:17}
can be performed using the Fourier series representation of the exponential
function \cite{gra80}. After straightforward integration we obtain
\begin{eqnarray}
&&H\left( \mathbf{k},\omega ,t\right) =\sum_{n,l=-\infty }^{\infty
}J_{n}\left( k_{\perp }a\right) e^{i\left( n-l\right) \psi }\left\{ \frac{%
-k_{\parallel }^{2}\Omega _{c}}{\left[ \omega -\zeta _{l}\left( \mathbf{k}%
\right) -i0\right] ^{2}}J_{l}\left( k_{\perp }a\right) \right.  \label{eq:19} \\
&&\left. +\frac{k_{\perp }^{2}}{2}\frac{1}{\omega -\zeta _{l}\left( \mathbf{k%
}\right) -i0}\left[ J_{l+1}\left( k_{\perp }a\right) e^{-i\psi
}-J_{l-1}\left( k_{\perp }a\right) e^{i\psi }\right] \right\} .  \nonumber
\end{eqnarray}%
Here $\psi =\varsigma (\theta -\varphi )+\Omega _{c}t$, $\varphi $ is the initial
phase of the cyclotron motion of the test particle with $v_{0x}=v_{0\perp }\cos \varphi $ and
$v_{0y}=v_{0\perp }\sin \varphi $, $\theta $ is the azimuthal angle of the transverse wave vector
$\mathbf{k}_{\perp }$ with $k_{x}=k_{\perp }\cos \theta $ and $k_{y}=k_{\perp }\sin
\theta $. Next we substitute Eq.~\eqref{eq:19} into Eq.~\eqref{eq:16} and
average the energy loss rate $dW_{\mathrm{st}}/dt$ with respect to the particle
cyclotron period $2\pi /\Omega _{c}$. Then in the term containing $k_{\parallel }^{2}$
performing integrations by parts in the $\omega $-integral, one finally obtains
\begin{eqnarray}
&&S_{\mathrm{st}} \equiv \left\langle \dfrac{dW_{\mathrm{st}}}{dt}%
\right\rangle =\frac{\pi Z^{2}e^{2}}{M}\int \frac{d\mathbf{k}}{k^{2}}%
\int_{-\infty }^{\infty }d\omega \sum_{l=-\infty }^{\infty }\delta \left(
\omega -\zeta _{l}\left( \mathbf{k}\right) \right)  \label{eq:20} \\
&&\times \left\{ k_{\parallel }^{2}J_{l}^{2}\left( k_{\perp }a\right) \frac{%
\partial }{\partial \omega }\left[ \omega P\left( \mathbf{k},\omega \right) %
\right] +\frac{k_{\perp }^{2}}{2\Omega _{c}}\left[ J_{l-1}^{2}\left(
k_{\perp }a\right) -J_{l+1}^{2}\left( k_{\perp }a\right) \right] \left[
\omega P\left( \mathbf{k},\omega \right) \right] \right\} .  \nonumber
\end{eqnarray}%
The above expression is the main result of this paper. For a simplicity we
shall call the expression \eqref{eq:20} as a Bessel-function representation
of the stochastic energy loss rate. For many practical applications, however, it
is important to represent the energy loss in an alternative but equivalent
integral form, see Appendix~\ref{sec:app1} for details. In particular, the
integral representation of the energy loss facilitates the limits of a heavy
ion (with $M\gg m$, where $m$ is the electron mass) and a weak magnetic
field when $\Omega _{c}$ is vanishingly small. Although the stochastic part
of the energy loss, Eq.~\eqref{eq:20}, vanishes in the limit $M\to \infty $
for pedagogical purposes we treat this case first keeping the
leading term only (proportional to $M^{-1}$). In this case the projectile
moves with rectilinear trajectory with an arbitrary angle with respect to
the magnetic field. The limit $\Omega _{c}\to 0$ of $S_{\mathrm{st}}$ is
performed in Appendix~\ref{sec:app1}, see Eq.~\eqref{eq:a3}. The
second term of this expression contains a full derivative and vanishes after
$\omega $-integration and the stochastic energy loss is then given by
\begin{equation}
S_{\mathrm{st}}=\frac{\pi Z^{2}e^{2}}{M}\int_{0}^{2\pi }\frac{d\varphi }{%
2\pi }\int d\mathbf{k}\left\{ \frac{\partial }{\partial \omega }\left[
\omega P\left( \mathbf{k},\omega \right) \right] \right\} _{\omega =\mathbf{k%
}\cdot \mathbf{v}_{0}} .
\label{eq:21}
\end{equation}%
It should be noted that in Eq.~\eqref{eq:21} the effect of a magnetic field
is contained in the power spectrum $P(\mathbf{k},\omega )$ of
the fluctuations. It is seen that formally Eq.~\eqref{eq:21} is equivalent
to the stochastic energy loss derived in Refs.~\cite{bek66,sit67,akh75} at
vanishing magnetic field assuming that in Eq.~\eqref{eq:21} $P(\mathbf{%
k},\omega )$ is the power spectrum at $B=0$ and neglecting the phase
average $\int_{0}^{2\pi }d\varphi /2\pi $. In the present context of the
magnetized plasma this integration appears due to the average of the energy
loss \eqref{eq:20} with respect to the cyclotron period $2\pi /\Omega _{c}$
of the particle. Therefore for a conformity of our results with the case of
unmagnetized plasma at $B\to 0$ the $\varphi $-average is
unnecessary and must be skipped.

It is well known that un upper cutoff parameter $k_{\max }=1/r_{\min }$
(where $r_{\min }$ is the effective minimum impact parameter) must be
introduced in Eqs.~\eqref{eq:14}, \eqref{eq:20} and \eqref{eq:21} to
avoid the logarithmic divergence at large $k$. This divergence corresponds
to the incapability of the perturbation theory to treat close encounters
between the projectile particle and the plasma electrons properly. For $r_{\min }$
we use the effective minimum impact parameter excluding hard Coulomb
collisions with a scattering angle larger than $\pi /2$. The resulting
cutoff parameter $k_{\max }$ is well known for energy loss calculations
(see, e.g., Refs.~\cite{zwi99,ner07} and references therein) and reads $%
k_{\max }=m(v_{0}^{2}+v_{\mathrm{th}}^{2})/|Z|e^{2}$, where $v_{\mathrm{th}%
}=(T/m)^{1/2}$ is the thermal velocity of the electrons.

Within linear response theory, the power spectrum $P(\mathbf{k},\omega )$ of
the fluctuations of the electrostatic fields at thermal equilibrium follows
from the fluctuation-dissipation theorem and is completely determined by the
longitudinal dielectric function $\varepsilon (\mathbf{k},\omega )$ of the
medium \cite{sit67,akh75},
\begin{equation}
P\left( \mathbf{k},\omega \right) =\frac{4T}{\left( 2\pi \right) ^{3}\omega }%
\mathrm{Im}\frac{-1}{\varepsilon \left( \mathbf{k},\omega \right) }.
\label{eq:22}
\end{equation}%
We consider magnetized electron plasma when the longitudinal dielectric
function is given by (see, e.g., Ref.~\cite{ich73})
\begin{equation}
\varepsilon \left( \mathbf{k},\omega \right) =1+\frac{1}{k^{2}\lambda _{%
\mathrm{D}}^{2}}\left\{ 1+\sum_{n=-\infty }^{\infty }\frac{\omega }{\omega
-n\omega _{c}}\Lambda _{n}\left( z\right) \left[ W\left( \beta _{n}\right) -1%
\right] \right\} ,  \label{eq:23}
\end{equation}%
where $z=k_{\perp }^{2}a_{e}^{2}$, $\beta _{n}=(\omega -n\omega_{c})/|k_{\parallel }|v_{\mathrm{th}}$,
$\Lambda _{n}(z)=e^{-z}I_{n}(z)$, $I_{n}$ is the modified Bessel function of the first kind, $a_{e}=%
v_{\mathrm{th}}/\omega_{c}$ is the cyclotron radius of the electrons. Also $W(\zeta )=g(\zeta )+if(\zeta )$
is the plasma dispersion function with the real ($g(\zeta )$) and imaginary ($f(\zeta )$) parts \cite{fri61}
\begin{equation}
g\left( \zeta \right) =1+\frac{\zeta }{\sqrt{2\pi }}\int_{-\infty }^{\infty }%
\frac{e^{-t^{2}/2}dt}{t-\zeta },\quad f\left( \zeta \right) =\sqrt{\frac{\pi
}{2}}\zeta e^{-\zeta ^{2}/2} .
\label{eq:24}
\end{equation}

In the next sections we consider some particular cases of the stochastic energy loss of the projectile
particle, in particular at vanishing and infinitely strong magnetic fields. At vanishing magnetic field
we denote the dielectric function as $\varepsilon _{0}(k,\omega )$ which is isotropic and is given by the
usual expression (see, e.g., Ref.~\cite{ich73})
\begin{equation}
\varepsilon _{0}\left(k,\omega \right) =1+\frac{1}{k^{2}\lambda _{%
\mathrm{D}}^{2}}W\left( \frac{\omega }{kv_{\mathrm{th}}}\right) .
\label{eq:25}
\end{equation}%
To the dielectric function \eqref{eq:25} corresponds the isotropic power spectrum $P_{0}(k,\omega )$ according
to Eq.~\eqref{eq:22}. In this case the stochastic energy loss rate \eqref{eq:21} is simplified to \cite{bek66,sit67,akh75}
\begin{equation}
S_{\mathrm{st}}=\frac{(2\pi )^{2}Z^{2}e^{2}}{M} \int_{0}^{k_{\max}} P_{0}\left(k,kv_{0}\right) k^{2} dk .
\label{eq:xxx}
\end{equation}%

At infinitely strong magnetic field we denote the dielectric function as $\varepsilon _{\infty }(\mathbf{k},\omega )$
which is determined by Eq.~\eqref{eq:25}, where, however, the phase velocity $\omega /k$ in the argument of the plasma
dispersion function is replaced by $\omega /|k_{\parallel }|$ (see, e.g., Ref.~\cite{ner07}).

The above expressions gives the mean energy (per unit time) absorbed (or emitted) by a propagating projectile particle
from the heat reservoir. Physically, this arises from the energy absorption from thermal fluctuations. In the absence
of the magnetic field it is to be noted here that because the spectral density $P_{0}(k,\omega )$ of the fluctuations
of the electric fields is positive for positive frequencies by definition, according to Eq.~\eqref{eq:xxx} the particle
energy will grow due to interactions with the fluctuating electric fields. The present situation with a magnetized
plasma may be quit different. It will be further discussed in Sec.~\ref{sec:3}.

\section{ELR without and with strong magnetic field}
\label{sec:3}

To illustrate the problem of the stochastic energy loss and the interrelation of the treatments determined
by Eqs.~\eqref{eq:14} and \eqref{eq:20}, we consider the cases without and with an infinitely strong magnetic
field. In the absence of a magnetic field the interaction of the test particle with stochastic plasma electric
fields yields a positive ELR (which corresponds to an energy gain by the particle) and the total ELR decreases.
At strong magnetic field, however, the stochastic interaction of the projectile may result in positive or negative
ELR depending on the velocity of the particle.

\subsection{ELRs at vanishing magnetic field}
\label{sec:3.1}

For a vanishing magnetic field when $a_{e}\gg \lambda_{\mathrm{D}}$ (or $\omega_{c}\ll \omega_{p}$), the power
spectrum of the fluctuation of the plasma electric fields is determined by Eq.~\eqref{eq:22}, where the longitudinal
dielectric function is given by the field-free expression~\eqref{eq:25}. At vanishing magnetic field the ELRs $S_{0}$
and $S_{\mathrm{st}}$ have been evaluated previously in Refs.~\cite{deu86,pet91} and \cite{bek66,sit67,akh75}
(see also references therein), respectively. In the leading order with respect to the cutoff parameter $k_{\max }$
the ELRs $S_{0}$ and $S_{\mathrm{st}}$ read
\begin{eqnarray}
&&S_{0}=-Q_{0}\frac{\pi }{\lambda }\left\{ \ln \xi \left[ \mathrm{erf}\left(
\frac{\lambda }{\sqrt{2}}\right) -\frac{2}{\pi }f\left( \lambda \right) %
\right] -\frac{2}{\pi }\int_{0}^{\lambda }\Xi \left( u\right) udu\right\} , \label{eq:26} \\
&&S_{\mathrm{st}}=Q_{0}\frac{m}{M}\frac{2}{\lambda }\left[ f\left( \lambda
\right) \ln \xi -\Xi \left( \lambda \right) \right] , \label{eq:27}
\end{eqnarray}%
where $\mathrm{erf}(z)$ is the error function, $Q_{0}=4n_{e}Z^{2}e^{4}/mv_{%
\mathrm{th}}$, $\xi =k_{\max }\lambda _{\mathrm{D}}=(\lambda
^{2}+1)/\epsilon |Z|$, $\lambda =v_{0}/v_{\mathrm{th}}$ is the projectile
velocity in units of thermal velocity. Also
\begin{equation}
\Xi \left( u\right) =\frac{1}{4}f\left( u\right) \ln \left[ g^{2}\left(
u\right) +f^{2}\left( u\right) \right] +\frac{1}{2}g\left( u\right) \left[
\frac{\pi }{2}-\arctan \frac{g\left( u\right) }{f\left( u\right) }\right] .
\label{eq:28}
\end{equation}%
It is well known that at vanishing magnetic field the ELRs determined by $%
S_{0}$ and $S_{\mathrm{st}}$ are always negative and positive, respectively.
This features correspond to the energy loss and energy gain, respectively,
by projectile particle.

Consider briefly some limiting cases for the low- and high-velocity
projectiles. Taking into account the behavior of the function $\Xi (u)\simeq
(1/2)\sqrt{\pi /2}u$ (at $u\ll 1$) and $\Xi (u)\simeq -\pi /2u^{2}$ (at $%
u\gg 1$), at low- ($\lambda \ll 1$) and high-velocities ($\lambda \gg 1$) the ELRs become
\begin{eqnarray}
&&S_{0}\simeq -Q_{0}\frac{\sqrt{2\pi }}{3}\lambda ^{2}\left( \ln \frac{1}{%
\epsilon |Z|}-\frac{1}{2}\right) ,\quad S_{\mathrm{st}}\simeq Q_{0}\frac{m}{M%
}\sqrt{2\pi }\left( \ln \frac{1}{\epsilon |Z|}-\frac{1}{2}\right) , \label{eq:29} \\
&&S_{0}\simeq -Q_{0}\frac{\pi }{\lambda } \ln \left( \frac{\lambda ^{3}}{%
\epsilon |Z|}\right) ,   \quad  S_{\mathrm{st}}\simeq Q_{0}\frac{m}{M}%
\frac{\pi }{\lambda ^{3}} , \label{eq:30}
\end{eqnarray}%
respectively. For derivation of the first relation in Eq.~\eqref{eq:30} we have used the frequency sum rule for
the dielectric function~\eqref{eq:25}. From Eqs.~\eqref{eq:29} and \eqref{eq:30} it is seen that for light
projectile particle (for instance for an electron with $M=m$) the stochastic ELR may essentially contribute to
the total energy loss at small velocities. Then the total rate $S_{0}+S_{\mathrm{st}}$ becomes positive which
corresponds to an energy gain by projectile particle. At high-velocities, however, the ELR $S_{0}$ essentially
exceeds the stochastic energy loss and the total ELR remains negative.

\subsection{ELRs at infinitely strong magnetic field}
\label{sec:3.2}

We now turn to the case when a projectile particle moves in a plasma with a strong magnetic field, which is, on one
hand, sufficiently weak to allow a classical description ($\hbar\omega_{c}<T$), and, on the other hand, comparatively
strong so that the cyclotron frequency of the plasma electrons exceeds the plasma frequency $\omega_{c}\gg \omega_{p}$
(or $a_{e}\ll \lambda_{\mathrm{D}}$). This limits the values of the magnetic field, the temperature, and the plasma
density. From these conditions we can obtain $3\times 10^{-6}\sqrt{n_{e}} <B<10^{5}T$, where $n_{e}$ is measured in
cm$^{-3}$, $T$ in eV, and $B$ in kG. The plasma electrons in the lowest order can respond to the excited waves only
in the direction of $\mathbf{B}$. Because of this assumption, the perpendicular cyclotron motion of the test and plasma
particles is neglected. As shown by Rostoker and Rosenbluth \cite{ros60} the plasma waves involved in this case are
oblique plasma waves having the approximate dispersion relation $\omega =\omega_{p} k_{\parallel}/k$.

The dielectric function $\varepsilon _{\infty }(\mathbf{k},\omega )$ is obtained from Eq.~\eqref{eq:25},
where, however, the phase velocity $\omega /k$ in the argument of the plasma dispersion function is replaced
by $\omega /|k_{\parallel }|$. The corresponding power spectrum of the fluctuations in the presence of strong
magnetic field is connected to $\varepsilon _{\infty }(\mathbf{k},\omega )$ via relation~\eqref{eq:22}.
Neglecting the transversal cyclotron motion of the test particle and the electrons in Eqs.~\eqref{eq:14}
and \eqref{eq:20} in the lowest order one obtains (see, e.g., Ref.~\cite{ner98,ner07})
\begin{eqnarray}
&&S_{0}=-Q_{0}\lambda \left[ f\left( \lambda \right) \ln \xi -\Xi \left(
\lambda \right) \right] , \label{eq:32} \\
&&S_{\mathrm{st}}=\frac{\pi Z^{2}e^{2}}{M}\int d\mathbf{k}\frac{k_{\parallel
}^{2}}{k^{2}}\frac{\partial }{\partial \omega }\left[ \omega P\left( \mathbf{%
k},\omega \right) \right] _{\omega =k_{\parallel }v_{0\parallel }} , \label{eq:33}
\end{eqnarray}
where now $\lambda =v_{0\parallel }/v_{\mathrm{th}}$. Substituting the power spectrum of the fluctuations
with the dielectric function $\varepsilon _{\infty}(\mathbf{k},\omega )$ in Eq.~\eqref{eq:33} in the leading
order with respect to the cutoff parameter $\xi$ we arrive at
\begin{equation}
S_{\mathrm{st}}=Q_{0}\frac{m}{M}\frac{\partial }{\partial \lambda }\left[
f\left( \lambda \right) \ln \xi -\Xi \left( \lambda \right) \right] ,
\label{eq:34}
\end{equation}
where $\Xi (\lambda )$ is determined by Eq.~\eqref{eq:28}.

\begin{figure}[tbp]
\includegraphics[width=80mm]{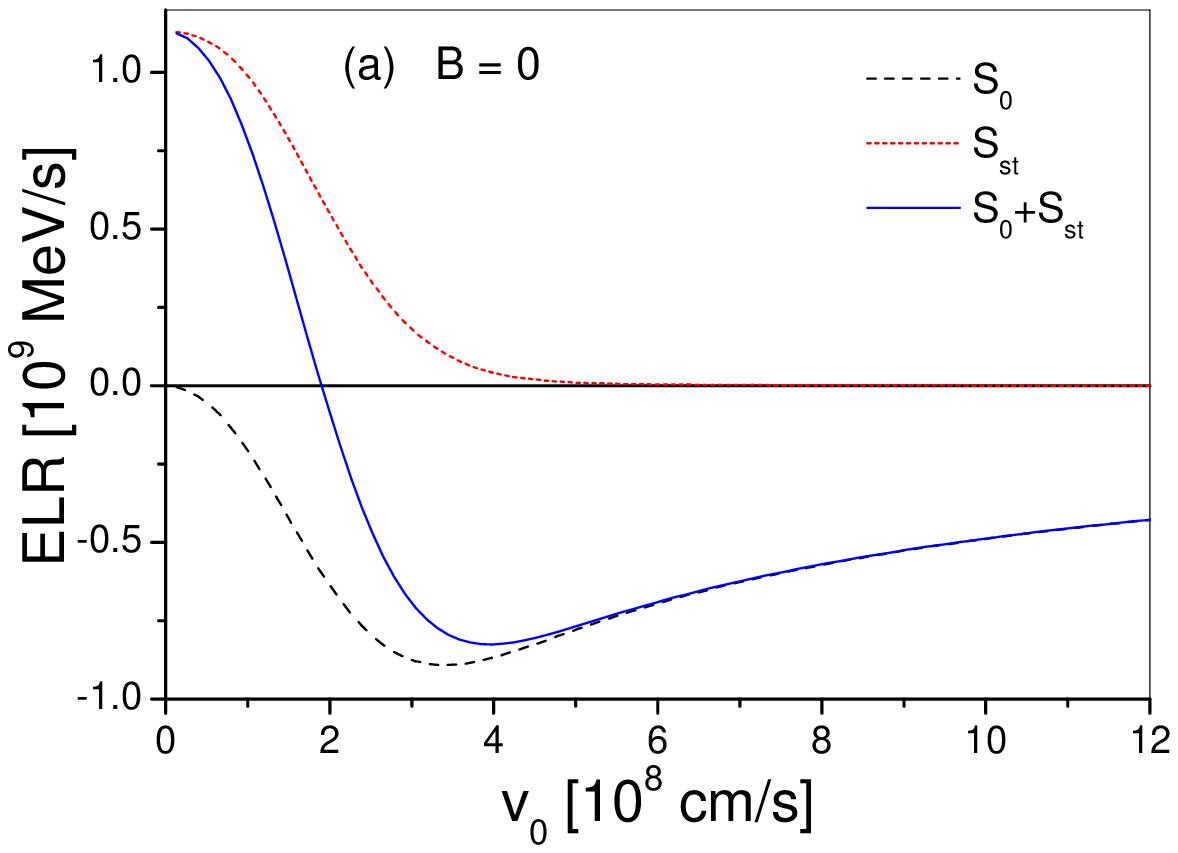}
\includegraphics[width=80mm]{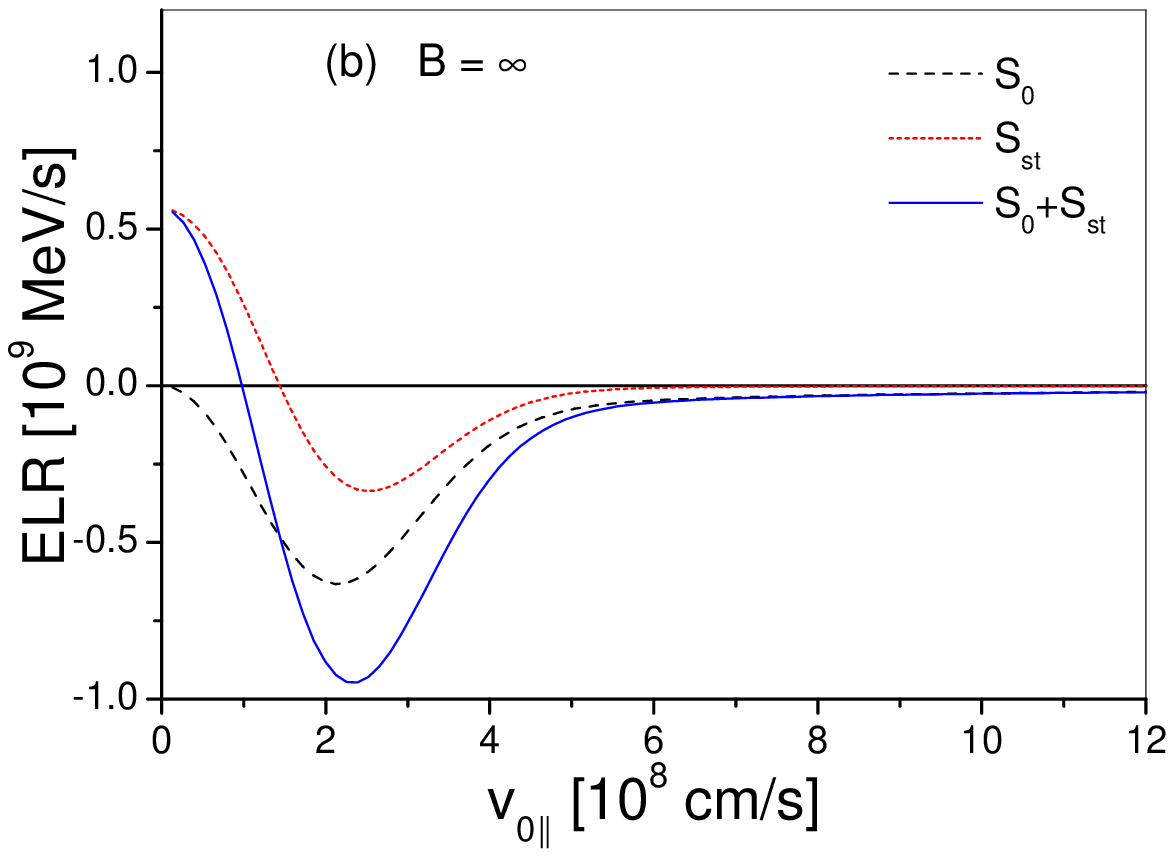}
\caption{(Color online) Electron energy loss rate (in units of $10^{9}$~MeV/s) in a plasma with
$n_{e}=10^{20}$~cm$^{-3}$, $T=10$~eV ($\epsilon =0.06$) in terms of the velocity (in units of
$10^{8}$~cm/s) at vanishing (a) and strong (b) magnetic field. The energy loss rates $S_{0}$ and
$S_{\mathrm{st}}$ and the total rate $S_{0}+S_{\mathrm{st}}$ are shown as dashed, dotted and solid
lines, respectively.}
\label{fig:1}
\end{figure}

Consider briefly the limits of low- and high-velocities. At $\lambda \ll 1$ from Eqs.~\eqref{eq:32} and
\eqref{eq:34} one obtains
\begin{eqnarray}
S_{0}\simeq -Q_{0}\sqrt{\frac{\pi }{2}}\lambda ^{2}\left( \ln \frac{1}{%
\epsilon |Z|}-\frac{1}{2}\right) , \label{eq:35} \\
S_{\mathrm{st}}\simeq Q_{0}\frac{m}{M}\sqrt{\frac{\pi }{2}}\left( \ln \frac{1%
}{\epsilon |Z|}-\frac{1}{2}\right) . \label{eq:36}
\end{eqnarray}

At high-velocities, $\lambda \gg 1$, Eqs.~\eqref{eq:32} and \eqref{eq:34} become
\begin{equation}
S_{0}\simeq -Q_{0}\frac{\pi }{2\lambda },\quad S_{\mathrm{st}}\simeq -Q_{0}%
\frac{m}{M}\frac{\pi }{\lambda ^{3}} ,
\label{eq:37}
\end{equation}
respectively. It is seen that unlike the field-free case where the energy loss rate $S_{\mathrm{st}}$ in
Eqs.~\eqref{eq:29} and \eqref{eq:30} is always positive, the ELR in the presence of a strong magnetic
field is positive at small-velocities while it is negative at high-velocities. For a light particle
(e.g. for an electron) the stochastic ELR even essentially exceeds the energy loss $S_{0}$ at small
velocity limit and the total ELR becomes positive. The results of the numerical evaluation of Eqs.~\eqref{eq:26},
\eqref{eq:27} and \eqref{eq:32}, \eqref{eq:34} are shown in Fig.~\ref{fig:1} for an electron test particle ($M=m$)
moving in a plasma with $n_{e}=10^{20}$~cm$^{-3}$ and $T=10$~eV ($\epsilon =0.06$).
This figure shows the ELR (in units of $10^{9}$~MeV/s) as a function of the particle velocity (in units
of $10^{8}$~cm/s) at vanishing (left panel) and at strong (right panel) magnetic field. It is evident that the effect of
the electric field fluctuations on the test particle energy loss is significant at low-velocities both
for vanishing and strong magnetic field. At higher velocities the relative importance of the stochastic
energy loss to the usual energy loss $S_{0}$ decreases gradually. We also mention that similar calculation
can be done for any test ion with $M\gg m$. However, the stochastic energy loss for an ion both at vanishing
and infinitely strong magnetic field at least three orders of magnitude smaller than $S_{\mathrm{st}}$ shown
in Fig.~\ref{fig:1}.

We would like to close this section with the following remark. From Eqs.~\eqref{eq:29} and \eqref{eq:36}
it is seen that the fluctuation energy loss rate does not vanish at low-velocities (as it occurs with the
polarization energy loss rates \eqref{eq:29} and \eqref{eq:35}) for vanishing and strong magnetic field
and the corresponding friction coefficient diverges as $v_{0}\to 0$. For arbitrary magnetic field similar
conclusion can be made on the basis of a general expression~\eqref{eq:a2} where the stochastic ELR is
given in the integral representation. At the vanishing ion velocity (i.e. at $v_{0}\to 0$) this expression
yields
\begin{equation}
S_{\mathrm{st}} =\frac{\pi Z^{2}e^{2}}{M}\int \frac{d\mathbf{k}}{k^{2}}%
\left[ k_{\parallel }^{2} P\left( \mathbf{k},0 \right) %
+k_{\perp }^{2} P\left( \mathbf{k},\Omega_{c} \right) \right] .
\label{eq:yyy}
\end{equation}
It is seen that in low--velocity limit and for arbitrary $B$ the rate $S_{\mathrm{st}}$ is finite and
its longitudinal and transversal parts are determined by static $P(\mathbf{k},0)$ and dynamic $P(\mathbf{k},%
\Omega_{c})$ fluctuation spectra, respectively. At vanishing magnetic field (i.e. at $\Omega_{c}\to 0$) the
second term in Eq.~\eqref{eq:yyy} can be collected with the first one and the resulting ELR coincides with
Eq.~\eqref{eq:xxx} in the limit $v_{0}\to 0$. In this case the stochastic energy loss is only determined by the
static fluctuation spectra $P(k,0)$. This feature of the stochastic ELR at vanishing velocity is closely resembled
the Brownian motion. In fact at $Mv^{2}/2 \lesssim T$ when the kinetic energy of the particle is comparable
with the thermal energy of the plasma environment the particle motion is stochastic in nature. The energy change
of the particle is determined by the correlation function $C(t)=\langle\mathbf{v}_{\mathrm{st}}(t)%
\cdot \mathbf{F}_{\mathrm{st}}(t)\rangle$, where $\mathbf{F}_{\mathrm{st}}(t)$ is the stochastic force acting
on the particle and $\mathbf{v}_{\mathrm{st}}(t)$ is the stochastic velocity of the particle under the influence of $\mathbf{F}_%
{\mathrm{st}}(t)$. At low velocities $\mathbf{v}_{\mathrm{st}}(t)$ can be estimated employing the Langevin equation
and the result at long times reads $\mathbf{v}_{\mathrm{st}}(t) \simeq (1/\gamma M) \mathbf{F}_{\mathrm{st}}(t)$, where
$\gamma$ is the friction coefficient. Thus, asymptotically $C(t)\simeq (1/\gamma M) \langle\mathbf{F}^{2}_{\mathrm{st}}(t)\rangle$
is a constant at thermal equilibrium which is determined by the thermodynamic properties of the medium. At vanishing
magnetic field this constant is related to the static fluctuation spectrum $P(k,0)$ while in $B\neq 0$ case the
fluctuations give rise the cyclotron motion of the particle and the transversal part of $C(t)$ is determined by the
dynamic spectrum $P(\mathbf{k},\Omega_{c})$, see the second term in Eq.~\eqref{eq:yyy}.

\section{ELR at low--velocities and for intermediate magnetic field}
\label{sec:4}

In Sec.~\ref{sec:3}, the effect of the electric--field fluctuations on the energy loss of the projectile particle has
been investigated in two extreme regimes with vanishing and very strong magnetic fields. In this section we extend
these results considering briefly some intermediate cases with finite magnetic field which cover the entire domain between
field--free and strong magnetic field regimes. As will be shown below at $a_{e}\sim \lambda_{\mathrm{D}}$ the effect of
the magnetic field on the energy loss process shows some additional features. In general, the analytical or numerical
investigation of the energy loss rates given by Eqs.~\eqref{eq:14} and \eqref{eq:20} with Eqs.~\eqref{eq:22} and \eqref{eq:23}
for arbitrary $B$ and the test particle velocity $v$ is a formidable task and requires separate investigations. Previously
the energy loss $S_{0}$ has been investigated numerically and analytically for arbitrary $B$ and $v$ but mostly neglecting
the gyration of the test particle, Refs.~\cite{ner07,mol03,cer00,ros60,akh61,hon63,may70,ner98,ner00,ner03,ner11,see98,wal00,ste01,deu08}.
To gain more insight into the magnetic field effect on the stochastic energy loss process we consider here the limit of
low--velocity projectile ion when electric field fluctuations are expected to be very important. In this case assuming
a heavy ion ($M\gg m$) the imaginary part of the inverse dielectric function (spectral function) involved in Eq.~\eqref{eq:22}
is represented in the form
\begin{equation}
\mathrm{Im}\frac{-1}{\varepsilon \left( \mathbf{k},\omega \right) }=\frac{%
\mathrm{Im}\left[ \varepsilon \left( \mathbf{k},\omega \right) \right] }{%
\left\vert \varepsilon \left( \mathbf{k},\omega \right) \right\vert ^{2}}%
\simeq \frac{\mathrm{Im}\left[ \varepsilon \left( \mathbf{k},\omega \right) %
\right] }{\left\vert \varepsilon \left( \mathbf{k},0\right) \right\vert ^{2}} .
\label{eq:n1}
\end{equation}
Note that in the last part of Eq.~\eqref{eq:n1} the dynamical screening has been replaced by the static one thus neglecting
the collective plasma resonances involved in $\varepsilon (\mathbf{k},\omega)$. For arbitrary velocities this simplified
version of the spectral function and the corresponding ELR has been calculated in Refs.~\cite{ner03,ner07} with vanishing
and strong magnetic fields. It has been shown that at $B=0$ the simplified and the full spectral functions yield close
ELRs at low--velocities. However, at $v_{0} >v_{\mathrm{th}}$ when the collective effects play an important role these
approaches deviate from each other. On the other hand, the strong magnetic field reduces the transversal dynamics of the
plasma electrons and the role of the collective effects becomes less important which results in the good agreement between
two approaches in a whole velocity range.

Consider the ELR $S_{0}$ given by Eq.~\eqref{eq:14} assuming that the projectile ion moves along the magnetic field (i.e
$v_{0\perp}=0$) with $v_{0\parallel}=v_{0} \ll v_{\mathrm{th}}$. Inserting the approximate spectral function \eqref{eq:n1}
into Eq.~\eqref{eq:14} and introducing new integration variables $\kappa =k^{2}a^{2}_{e}$, $\mu =\cos^{2}\theta$ (where
$\theta$ is the angle between $\mathbf{k}$ and $\mathbf{B}$) we arrive at
\begin{equation}
S_{0}\simeq -\sqrt{\frac{\pi }{8}}Q_{0}\lambda ^{2}\int_{0}^{\eta ^{2}\xi
^{2}}\frac{\kappa d\kappa}{\left(\kappa +\eta ^{2}\right) ^{2}}\int_{0}^{1}d\mu
\sum_{n=-\infty }^{\infty }\Lambda _{n}\left(\kappa \left( 1-\mu \right) \right)
\exp \left( -\frac{n^{2}}{2\kappa \mu }\right) ,
\label{eq:n8}
\end{equation}
where $\lambda =v_{0}/v_{\mathrm{th}}$, $\eta =a_{e}/\lambda _{\mathrm{D}}$, and the other quantities have been introduced
in Secs.~\ref{sec:2} and \ref{sec:3}.

The stochastic ELR at low--velocities is determined by Eq.~\eqref{eq:yyy} where the first and the second terms are the
contributions from longitudinal and transversal fluctuations, respectively. It should be emphasized that the relation
\eqref{eq:n1} is justified for the first term of Eq.~\eqref{eq:yyy} where only the static spectrum $P(\mathbf{k},0)$
is involved. The second term contains the cyclotron frequency $\Omega_{c}$ of the projectile ion. Since $\Omega_{c}\ll \omega_{c}$
the ion cannot resonate with plasma electron resonances involved in the dielectric function and in the fluctuation spectrum
$P(\mathbf{k},\Omega_{c})$ the simplification given by Eq.~\eqref{eq:n1} is again justified.

Consider now Eq.~\eqref{eq:yyy} with Eqs.~\eqref{eq:22} and \eqref{eq:n1}. Using the integration variables introduced
above we obtain
\begin{eqnarray}
&&S_{\mathrm{st}} \simeq \sqrt{\frac{\pi }{8}}\frac{m}{M}%
Q_{0}\int_{0}^{\eta ^{2}\xi ^{2}}\frac{\kappa d\kappa }{\left( \kappa +\eta
^{2}\right) ^{2}}\int_{0}^{1}\frac{d\mu }{\mu }\sum_{n=-\infty }^{\infty
}\Lambda _{n}\left( \kappa \left( 1-\mu \right) \right)   \label{eq:n2}  \\
&&\times \left\{ \mu \exp \left( -\frac{n^{2}}{2\kappa \mu }\right) +\left(
1-\mu \right) \exp \left[ -\frac{\left( n-\alpha \right) ^{2}}{2\kappa \mu }%
\right] \right\} ,  \nonumber
\end{eqnarray}
where $\alpha =\vert Z\vert m/M\ll 1$. Let us recall that at low--velocities the ELR, $S_{\mathrm{st}}$, in the leading
order does not depend on the ion velocity while $S_{0}\sim \lambda^{2}$ (see Eq.~\eqref{eq:n8}). It is seen that the
term of Eq.~\eqref{eq:n2} with $n=0$ (no cyclotron motion) exhibits a singular behavior in the limit of $\alpha \ll 1$
where the $\mu$ integration diverges logarithmically for small $\mu $ (or small $k_{\parallel}$). We must therefore keep
$\alpha $ finite in that integration to avoid such a divergence. This anomalous contribution that arises from the term
of Eq.~\eqref{eq:n2} with $n=0$ is denoted as $S_{\mathrm{st}}^{\prime }$. The other terms of Eq.~\eqref{eq:n2} with
$n\neq 0$ do not exhibit singular behavior at $\alpha \ll 1$ and their contribution to the stochastic energy loss is
denoted as $S_{\mathrm{st}}^{\prime \prime }$ ($S_{\mathrm{st}}=S_{\mathrm{st}}^{\prime }+S_{\mathrm{st}}^{\prime \prime }$).
Thus in the limit of a heavy projectile ion with $\alpha \ll 1$ from Eq.~\eqref{eq:n2} in the leading order we obtain
\begin{eqnarray}
&&S_{\mathrm{st}}^{\prime } = \sqrt{\frac{\pi }{8}}\frac{m}{M}%
Q_{0}\int_{0}^{\eta ^{2}\xi ^{2}}\frac{\kappa d\kappa }{\left( \kappa +\eta
^{2}\right) ^{2}}\left\{ \Lambda _{0}\left( \kappa \right)
\left(\ln\frac{2\kappa}{\alpha^{2}} -\gamma\right) \right.   \label{eq:n3}   \\
&&\left. +\int_{0}^{1}\left[ \Lambda _{0}\left( \kappa
\left( 1-\mu \right) \right) -\Lambda _{0}\left( \kappa \right) \right]
\frac{d\mu }{\mu }\right\} ,  \nonumber  \\
&&S_{\mathrm{st}}^{\prime \prime } =\sqrt{\frac{\pi }{2}}\frac{m}{M}%
Q_{0}\int_{0}^{\eta ^{2}\xi ^{2}}\frac{\kappa d\kappa }{\left( \kappa +\eta
^{2}\right) ^{2}}\int_{0}^{1}\frac{d\mu }{\mu }
\sum_{n=1}^{\infty }\Lambda _{n}\left( \kappa \left( 1-\mu
\right) \right) \exp \left( -\frac{n^{2}}{2\kappa \mu }\right) ,  \label{eq:n4}
\end{eqnarray}
where $\gamma =0.5772\ldots $ is the Euler's constant. The terms neglected in deriving Eqs.~\eqref{eq:n3} and
\eqref{eq:n4} are of the order of $\mathrm{O}(\alpha^{2})$ and $\mathrm{O} (\alpha^{2}\ln \alpha )$. Also we note that
the anomalous ELR $S_{\mathrm{st}}^{\prime }$ given by Eq.~\eqref{eq:n3} is valid at $\alpha <\eta $ (or $\Omega_{c}< \omega_{p}$)
which is easily fulfilled for heavy ions. In the opposite case with $\alpha >\eta $ (but $\alpha\ll 1$) one should consider
more general Eq.~\eqref{eq:n2} as a starting point for investigation of the anomalous ELR.

The ELR \eqref{eq:n3} can be further simplified assuming that $\eta >1/\xi $ (or $a_{e}>r_{\min }$). This condition
allows one to extend the upper limit of the integration in Eq.~\eqref{eq:n3} to infinity which yields
\begin{equation}
S_{\mathrm{st}}^{\prime }\simeq \sqrt{\frac{\pi }{8}}\frac{m}{M}Q_{0}\left[
\left( \ln \frac{2}{\alpha ^{2}}-\gamma -1\right) G_{0}\left( \eta
^{2}\right) +G_{1}\left( \eta ^{2}\right) \right]
\label{eq:n5}
\end{equation}
where
\begin{eqnarray}
&&G_{0}\left( x\right) =\int_{0}^{\infty }\frac{\Lambda _{0}\left( \kappa
\right) \kappa d\kappa }{\left( \kappa +x\right) ^{2}}=e^{x}\left[ \left(
1+x\right) K_{0}\left( x\right) -xK_{1}\left( x\right) \right] ,  \label{eq:n6}  \\
&&G_{1}\left( x\right) =\int_{0}^{\infty }\ln \left( \kappa +x\right) \frac{%
\Lambda _{0}\left( \kappa \right) \kappa d\kappa }{\left( \kappa +x\right) ^{2}} .
\label{eq:n7}
\end{eqnarray}
Here $K_{0}(x)$ and $K_{1}(x)$ are the modified Bessel functions of the second kind. The integral involved in the
function $G_{0}(x)$ has been evaluated by Ichimaru (see Ref.~\cite{ich73} for details).

\begin{figure}[tbp]
\includegraphics[width=80mm]{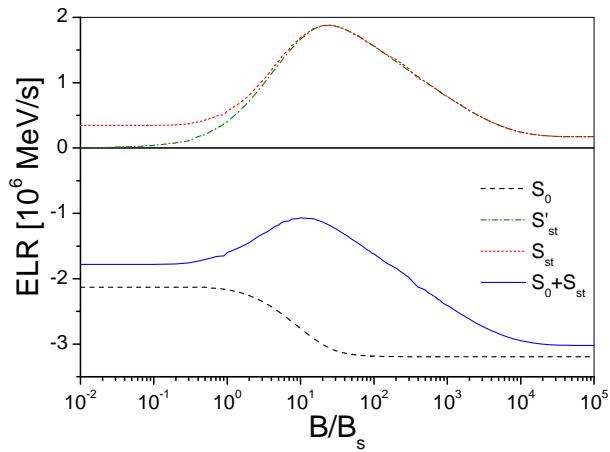}
\caption{(Color online) Proton energy loss rate (in units of $10^{6}$~MeV/s) in a plasma with
$n_{e}=10^{20}$~cm$^{-3}$, $T=10$~eV ($\epsilon =0.06$) vs scaled magnetic field $B/B_{s}$ and
at low--velocities ($\lambda =0.1$). The energy loss rates $S_{0}$, $S'_{\mathrm{st}}$
and $S_{\mathrm{st}}$ and the total rate $S_{0}+S_{\mathrm{st}}$ are shown as dashed, dot--dashed,
dotted and solid lines, respectively. See the text for explanations.}
\label{fig:2}
\end{figure}

Consider briefly the connection of Eqs.~\eqref{eq:n8}--\eqref{eq:n5} to the ELRs obtained in the previous section
for two limiting cases. It is seen from Eqs.~\eqref{eq:n8} and \eqref{eq:n2} that at very strong magnetic field
only the terms with $n=0$ contribute to the ELRs. Then the ELRs \eqref{eq:n8} and \eqref{eq:n2} in the limit of
$\eta \to 0$ after $\kappa$ integration coincide with Eqs.~\eqref{eq:35} and \eqref{eq:36}, respectively. We merely
note that in this limit the ELR $S_{\mathrm{st}}^{\prime\prime }$ vanishes and the stochastic energy loss in the
case of very strong magnetic field is determined by the anomalous term $S_{\mathrm{st}}^{\prime }$.

Next we treat the limit of the vanishing magnetic field ($\eta \to\infty$). In this case all cyclotron harmonics
in Eqs.~\eqref{eq:n8} and \eqref{eq:n2} contribute to the ELRs. It is useful to transform the $n$ summations
in these expressions into an integral form where the limit $\eta \to\infty$ can be easily done. Since this procedure
is similar to one considered in Appendix~\ref{sec:app1} we refer a reader to Refs.~\cite{ner07,ner11} for more
details. As a result we obtain that at vanishing magnetic field the $n$ summation in Eq.~\eqref{eq:n8} is replaced
by $\sqrt{\mu}$. The remaining integrals can be easily performed. After evaluating these integrals one can see that
the ELR $S_{0}$ at $\eta \to \infty$ completely agrees with the first relation in Eq.~\eqref{eq:29}. We turn now
to the anomalous term given by Eq.~\eqref{eq:n5}. Using the asymptotic behavior of the modified Bessel functions
one cane see that the functions $G_{0}(\eta^{2})$ and $G_{1}(\eta^{2})$ at $\eta \to \infty$ behave as $\sim 1/\eta$
and $\sim (1/\eta ) \ln\eta $, respectively, and thus vanish in the field--free case. On the other hand the stochastic
ELR $S_{\mathrm{st}}^{\prime\prime }$ is evaluated in a same manner as explained above and the summation in
Eq.~\eqref{eq:n4} in a limit of vanishing magnetic field yields $\sqrt{\mu}/2$. Substituting this formula into
Eq.~\eqref{eq:n4} and performing $\mu$ and $\kappa$ integrations we finally arrive at the second relation of
Eq.~\eqref{eq:29}. The anomalous term Eqs.~\eqref{eq:n3} and \eqref{eq:n5} therefore represent a new effect arising
from the presence of the magnetic field. The physical origin of such an anomalous ELR is the same as those obtained
in Ref.~\cite{ich73} and Refs.~\cite{ner07,ner00} for the transport coefficients and the stopping power of an ion, respectively.
It may be traced to the spiral motion of the electrons along the magnetic field lines. These electrons naturally tend
to couple strongly with long-wavelength fluctuations (i.e., small $k_{\parallel}$) along the magnetic field. In addition,
when such fluctuations are characterized by slow variation in time (i.e., small $\omega =\Omega_{c}$), the contact
time or the rate of energy exchange between the electrons and the fluctuations will be further enhanced. In a plasma,
such low--frequency fluctuations are provided by the slow projectile ion. The above coupling can therefore be an
efficient mechanism of energy exchange between the electrons and the projectile ion. In the limit of $\Omega_{c} \to 0$
(or $M\to \infty$), the contact time thus becomes infinite and the ELR diverges.

In the presence of strong magnetic field one could expect that for heavy ions the anomalous ELR $S^{\prime}_%
{\mathrm{st}}$ mainly contributes to the total stochastic ELR $S_{\mathrm{st}}$. To demonstrate this feature the ELRs are
plotted in Fig.~\ref{fig:2} vs scaled magnetic field $B/B_{s}$ for $v_{0}= 0.1v_{\mathrm{th}} \simeq 1.3\times 10^{7}$~cm/s,
$n_{e}=10^{20}$~cm$^{-3}$, $T=10$~eV, where $B_{s}=mc\omega_{p}/e \simeq 3.2\times 10^{4}$~kG. The energy loss rates
$S_{0}$, $S'_{\mathrm{st}}$ and $S_{\mathrm{st}}$ and the total rate $S_{0}+S_{\mathrm{st}}$ are found for a proton
projectile and are shown as dashed, dot--dashed, dotted and solid lines, respectively. In Fig.~\ref{fig:2} the limiting
cases of the vanishing and very strong magnetic fields considered in Sec.~\ref{sec:3} are reached at $B\lesssim B_{s}$
and $B\gtrsim 10^{4}B_{s}$, respectively. It is seen that the magnetic field may essentially increase the ELRs in
plasma compared to the field--free (Eq.~\eqref{eq:29}) and strong field (Eqs.~\eqref{eq:35} and \eqref{eq:36}) regimes
and this occurs at $B\gtrsim 10B_{s}$. In the latter case the stochastic ELR is mainly determined by the anomalous
ELR $S'_{\mathrm{st}}$. Furthermore, at low ion velocities the fluctuations in the electric field essentially counter
the effect of the polarization field resulting in the reduced ELR (solid line in Fig.~\ref{fig:2}).

As discussed above, Eqs.~\eqref{eq:n8} and \eqref{eq:n2} and hence the results shown in Fig.~\ref{fig:2} account for
only static screening neglecting dynamic polarization effects. As shown in Refs.~\cite{ner07,ner03} in a vanishing
or very strong magnetic field these effects are not important at low--velocities, and the situation with finite $B$
requires further investigations, in particular for electron projectiles. Intuitively it is clear that a gyrating
test electron may resonate with a electron plasma fluctuations with a cyclotron frequency $\omega_{c}$, and the
approximation \eqref{eq:n1} is not as obvious as for the case of a heavy ion. Also it should be emphasized that the
validity of the regime of a classically strong magnetic field requires the condition $B<B_{c}=(mc/e\hslash ) T$.
Thus the results shown in Fig.~\ref{fig:2} are valid up to $B_{c}/B_{s}=T/\hslash\omega_{p} \simeq 27$. Clearly the
realization of the regime of a strong magnetic field requires high temperatures and low densities and the enhancement
of the ELRs at $B\gtrsim 10B_{s}$ may not be accessible under certain conditions.

\section{Summary}
\label{sec:disc}

In this paper within linear response theory we have investigated the energy loss rate (ELR) of a nonrelativistic
projectile particle moving
in a magnetized classical plasma due to the electric field fluctuations. In the course of this study we have derived
general expression for the stochastic energy loss rate. As in the field--free case the later is completely determined
by the fluctuation power spectrum $P(\mathbf{k},\omega)$. Assuming an equilibrated and homogeneous magnetized plasma
when the power spectrum of the fluctuations is derived from the fluctuation--dissipation theorem and is expressed via
longitudinal dielectric function of the medium we have considered two somewhat distinct cases with vanishing and
with extremely strong magnetic field. At vanishing magnetic field the electric field fluctuations lead to an energy
gain of the charged particle for all velocities. Physically, this is arisen due to the energy absorption by projectile
particle from thermal fluctuations. It has been shown that in the presence of strong magnetic field this effect occurs
only at low--velocities. At high--velocities the test particle systematically loses its energy due to the interaction
with a stochastic electric field.

In the low--velocity limit but for arbitrary magnetic field, we have found an enhanced stochastic ELR of an ion compared
to the limiting cases with vanishing and very strong magnetic fields, mainly due to the strong coupling between the
spiral motion of the electrons and the long--wavelength, low--frequency fluctuations excited by the projectile. The
nature of this anomalous ELR is only conditioned by the external magnetic field.

The analysis presented above can in principle be extended to the case of a nonequilibrium magnetized plasma, provided
the power spectrum of the electrostatic fluctuations are known. In principle the power spectrum can be derived within
linear response theory \cite{sit67,akh75}. Fluctuations are much stronger in nonequilibrium situations than in
systems in thermal equilibrium \cite{sit67,akh75}. The effect of the electric field fluctuations on the particle energy
loss is therefore expected to be stronger in a nonequilibrium plasma. We intend to address this issue in our forthcoming
investigations.

\begin{acknowledgments}
We gratefully acknowledge Prof. C. Toepffer, Dr. G. Zwicknagel and Dr. Amal K. Das for many fruitful discussions. The work
of H.B.N. has been partially supported by the State Committee of Science of Armenian Ministry of Higher Education and Science
(Project No.~11-1c317).
\end{acknowledgments}

\appendix

\section{Integral representation of the stochastic energy loss}
\label{sec:app1}

In Sec.~\ref{sec:2} we have derived the Bessel-function representation of
the stochastic energy loss of a gyrating test particle. For some applications an
integral representation of $S_{\mathrm{st}}$ is desirable. For deriving the
integral form of the energy loss we rewrite the delta-function $\delta
(\omega -\zeta _{l}(\mathbf{k}))$ in Eq.~\eqref{eq:20} using an integral
representation of this function. Then Eq.~\eqref{eq:20} becomes
\begin{eqnarray}
&&S_{\mathrm{st}} =\frac{Z^{2}e^{2}}{2M}\int \frac{d\mathbf{k}}{k^{2}}%
\int_{-\infty }^{\infty }d\omega \int_{-\infty }^{\infty }dte^{i(\omega
-k_{\parallel }v_{0\parallel })t}\sum_{l=-\infty }^{\infty }e^{-il\Omega
_{c}t}J_{l}^{2}\left( k_{\perp }a\right)  \label{eq:a1} \\
&&\times \left\{ k_{\parallel }^{2}\frac{\partial }{\partial \omega }\left[
\omega P\left( \mathbf{k},\omega \right) \right] -ik_{\perp }^{2}\frac{\sin
\left( \Omega _{c}t\right) }{\Omega _{c}}\left[ \omega P\left( \mathbf{k}%
,\omega \right) \right] \right\} .  \nonumber
\end{eqnarray}%
Then using the summation formula $\sum_{l=-\infty }^{\infty
}e^{ilt}J_{l}^{2}(z)=J_{0}(2z\sin \frac{t}{2})$ \cite{gra80} the stochastic
energy loss may be alternatively represented in the form
\begin{eqnarray}
&&S_{\mathrm{st}} =\frac{Z^{2}e^{2}}{M}\int \frac{d\mathbf{k}}{k^{2}}%
\int_{-\infty }^{\infty }d\omega \int_{0}^{\infty }J_{0}\left( 2k_{\perp
}a\sin \frac{\Omega _{c}t}{2}\right) dt  \label{eq:a2} \\
&&\times \left\{ k_{\parallel }^{2}\frac{\partial }{\partial \omega }\left[
\omega P\left( \mathbf{k},\omega \right) \right] \cos \left[ \left( \omega
-k_{\parallel }v_{0\parallel }\right) t\right] +k_{\perp }^{2}\left[ \omega
P\left( \mathbf{k},\omega \right) \right] \frac{\sin \left( \Omega
_{c}t\right) }{\Omega _{c}}\sin \left[ \left( \omega -k_{\parallel
}v_{0\parallel }\right) t\right] \right\} .  \nonumber
\end{eqnarray}%
Using this integral form it is now easy to perform the limit of a heavy ion
(with $M\to \infty $) when $\Omega _{c}\rightarrow 0$. In this limit
the zero order Bessel function becomes $J_{0}(k_{\perp }v_{0\perp }t)$ which
we represent in the integral form as an average of the exponential function $%
e^{\pm i\mathbf{k}_{\perp }\cdot \mathbf{v}_{0\perp }t}$ with respect to the
azimuthal angle $\varphi $ of the vector $\mathbf{v}_{0\perp }$. Thus performing
$t$-integration the energy loss \eqref{eq:a2} in the limit $\Omega _{c}\to 0$ becomes
\begin{eqnarray}
&&S_{\mathrm{st}} =\frac{\pi Z^{2}e^{2}}{M}\int_{0}^{2\pi }\frac{d\varphi }{%
2\pi }\int d\mathbf{k}\int_{-\infty }^{\infty }d\omega  \label{eq:a3} \\
&&\times \left\{ \delta \left( \omega -\mathbf{k}\cdot \mathbf{v}_{0}\right)
\frac{\partial }{\partial \omega }\left[ \omega P\left( \mathbf{k},\omega
\right) \right] -\frac{k_{\perp }^{2}}{k^{2}}\frac{\partial }{\partial
\omega }\left[ \delta \left( \omega -\mathbf{k}\cdot \mathbf{v}_{0}\right)
\omega P\left( \mathbf{k},\omega \right) \right] \right\} .  \nonumber
\end{eqnarray}%
The second term in Eq.~\eqref{eq:a3} contains full derivative and vanishes
after $\omega $-integration. Then the averaged stochastic energy loss is
solely given by the first term, see Eq.~\eqref{eq:21}.

\end{document}